\documentclass[a4paper,fleqn]{cas-sc}

\usepackage[authoryear]{natbib}
\usepackage{array}

\def\tsc#1{\csdef{#1}{\textsc{\lowercase{#1}}\xspace}}
\tsc{WGM}
\tsc{QE}
\tsc{EP}
\tsc{PMS}
\tsc{BEC}
\tsc{DE}

\begin{document}
\let\WriteBookmarks\relax
\def\floatpagepagefraction{1}
\def\textpagefraction{.001}
\shorttitle{Sensitivity of ...}
\shortauthors{A. Saha}

\title [mode = title]{Sensitivity of $^{107,109}$Ag($\alpha$,xn) cross sections to statistical-model inputs}                      



\author[1]{Arunabha Saha}

\ead{arunabhaiitb@gmail.com}

\credit{Writing-original draft, visualization, software, methodology, formal analysis, conceptualization}

\affiliation[1]{organization={Department of Physics, ICFAI University Tripura},
                addressline={Kamalghat}, 
                city={Mohanpur},
                postcode={799210}, 
                state={Tripura},
                country={India}}


\begin{abstract}
The $\alpha$-induced reactions on silver isotopes leading to the production of the medically relevant radionuclides $^{108m,109g,110m,110g,111g}$In have been systematically analyzed using the TALYS~2.0 code. A total of 192 combinations of nuclear reaction model parameters---comprising level-density models (LDM), $\alpha$-optical model potentials ($\alpha$OMP), and pre-equilibrium (PE) models---were evaluated through $\chi^2$ minimization against the available experimental data.

The results reveal a pronounced channel-dependent sensitivity of the statistical-model ingredients. The $^{107}$Ag($\alpha$,3n)$^{108m}$In and $^{109}$Ag($\alpha$,3n)$^{110g}$In reactions are primarily governed by the level-density model. For the $^{107}$Ag($\alpha$,2n)$^{109g}$In reaction, the sensitivities to the LDM and the pre-equilibrium mechanism are comparable, indicating that both model ingredients play nearly equal roles in describing the experimental data. In contrast, the $^{107}$Ag($\alpha$,n)$^{110m}$In and $^{109}$Ag($\alpha$,2n)$^{111g}$In reactions are dominated by the PE mechanism, with the $\alpha$-optical model potential providing a secondary contribution and the LDM contributing only minimally. These findings demonstrate that the relative importance of the statistical-model ingredients varies significantly among the investigated reaction channels. Additionally, discrepancies between the present TALYS calculations and evaluated libraries such as TENDL-2023 are attributed to the absence of parameter optimization in the current approach.

Overall, the present analysis indicates that the relative importance of the statistical-model ingredients depends on the reaction channel and that no single parameter combination provides the best description of all investigated reactions. The observed channel-dependent sensitivities provide useful guidance for selecting and evaluating TALYS model ingredients for the studied reaction channels. Future work incorporating additional experimental datasets and model-uncertainty quantification would further assess the robustness of the preferred parameter combinations.

\end{abstract}


\begin{highlights}
\item TALYS-2.0 modeled $\alpha$-induced reactions on Ag producing medically relevant In radionuclides
\item $\chi^2$ based sensitivity analysis tested 192 model combinations systematically
\item Multi-neutron channels favor compound nucleus reactions
\item Reaction-specific models improve agreement with experimental cross sections
\end{highlights}


\maketitle

\section{Introduction}

In recent years, charged-particle-induced nuclear reactions have attracted significant attention within the nuclear data community due to their expanding applications in nuclear medicine, accelerator-based systems, and advanced nuclear technologies, as well as their importance in improving the theoretical understanding of reaction mechanisms. The field of nuclear medicine continues to evolve rapidly, driven by the development of novel radiopharmaceuticals and increasingly sophisticated instrumentation. Among these advancements, Positron Emission Tomography (PET) has emerged as a highly sensitive and quantitative functional imaging modality, offering wide-ranging applications in clinical diagnostics, treatment monitoring, and biomedical research~\citep{Kakavand2015}.

Accurate nuclear reaction cross-section modeling is essential for fusion reactor materials, shielding design, and medical isotope production. Previous studies have systematically investigated the role of nuclear level density models, gamma strength functions, and reaction mechanisms using codes such as TALYS~\citep{Koning2023}, EMPIRE~\citep{Herman2013}, and ALICE/ASH~\citep{Broeders2006}, demonstrating good harmony with experimental data for proton-, photon-, neutron-, and alpha-induced reactions across various isotopes~\citep{Ozdogan2018,Ozdogan2019,Kaplan2016}. In addition, advanced approaches, including Bayesian methods as well as artificial neural network (ANN) models trained using the Levenberg-Marquardt optimization algorithm, have demonstrated high predictive accuracy in estimating reaction cross-sections and level-density parameters~\citep{Ozdogan2021,Ozdogan2023,Ozdogan2024a}. These methodologies have been effectively applied to practical systems, including Ag isotopes relevant to fusion shielding and the production of medically important radionuclides such as $^{211}$At, $^{103}$Pd, and $^{125}$I~\citep{Ozdogan2024b,Uncu2023}, highlighting the importance of robust computational models validated against experimental data available in EXFOR~\citep{Otuka2014}.

Alpha-particle-induced reactions on silver isotopes produce many radionuclides that are extremely useful for medical and industrial applications. For example, the metastable isotope indium-110 ($^{110\text{m}}$In) has a half-life of 69.1 min and decays by positron emission ($E_{\beta^+}$ = 1011 keV, $I_{\beta^+}$ = 61.3\%). It is primarily a research radionuclide with potential for PET imaging, rather than one in routine clinical use. Additionally, $^{110\text{m}}$In emits a medium-energy, high-intensity gamma ray suitable for $\beta$+$\gamma$ coincidence PET~\citep{Sitarz2020}.

Similarly, the radionuclide $^{111}$In exhibits highly favorable nuclear decay characteristics for medical applications, undergoing electron-capture decay (T$_{1/2}$ = 2.80 days) to excited states of $^{111}$Cd, followed by cascade $\gamma$-ray emissions. The principal emissions include a 171.3 keV $\gamma$-ray (I$_\gamma$ = 90.3\%) and a subsequent 245.4 keV $\gamma$-ray (I$_\gamma$ = 94.0\%), arising from sequential de-excitation of intermediate excited states. It has been widely used in nuclear medicine for applications such as targeted diagnosis and treatment of liver tumors~\citep{Tang2019}, detection of gastric cancer and lymph-node metastases~\citep{Fujiwara2020}, imaging of colorectal cancer~\citep{DeGooyer2020}, and identification of myocardial damage~\citep{PONSLLADO20002198}, among others~\citep{Hermanne2014}. 

$^{111}$In can be produced through several nuclear reactions, including (1) alpha-particle-induced reactions on natural silver~\citep{TSOODOL2024111221}, (2) proton irradiation of tin~\citep{Hermanne2006}, and (3) proton irradiation of enriched $^{112}$Cd~\citep{GAO2021109828} or deuterium irradiation of natural cadmium~\citep{HERMANNE201619}. Among these, the $\alpha$-induced reaction on silver offers certain distinct advantages. In particular, the $^{109}$Ag($\alpha$,2n)$^{111}$In route can yield $^{111}$In with comparatively high radionuclidic purity due to the limited formation of long-lived contaminant isotopes. Additionally, natural silver consists of only two stable isotopes ($^{107}$Ag and $^{109}$Ag), simplifying target composition and reducing the complexity of by-product formation compared to multi-isotopic targets such as natural cadmium. This can ease radiochemical separation and improve product quality. Furthermore, the use of solid silver targets provides good thermal and mechanical stability under irradiation, making the $\alpha$-on-Ag pathway a valuable complementary production route, particularly when high-purity $^{111}$In is required.

The goal of this paper is to investigate the combination of level density model, alpha OMP, and pre-equilibrium model with comparatively improved agreement out of the 192 different combinations obtained using 6 different level density models, 8 different alpha optical model potentials, and 4 different pre-equilibrium models for each reaction using TALYS 2.0 code. In the present work, the results of theoretical calculation of cross section for each of the investigated reactions are compared with the experimental data taken from EXFOR library and TALYS Evaluated Nuclear Data Library (TENDL-2023) data~\citep{Koning2019}. The evaluated data of TENDL are based on parameter optimization within the TALYS framework. The objectives of the present work differ from those of TENDL, since this study investigates the predictive capability of the default TALYS-2.0 model combinations without parameter adjustment. For comparison with the experimental data, the value of reduced chi-square ($\chi^2$) (defined in Section~\ref{uses}) has been chosen as the guiding factor. Various other groups have reported similar type of research work~\citep{Canbula2024,Sekerci2022,Saha2026,Tureci2025}. For example, investigations on $^{92,94,95}$Mo isotopes using TALYS 1.96~\citep{Canbula2024} highlight the role of optical model potentials and nuclear level density models, where chi-square analysis enables optimal model selection with good agreement to EXFOR data. Similarly, in Ref.~\citep{Sekerci2022}, the effect of combination of level density model and alpha optical model potential on cross section estimation of $^{nat}$Sb($\alpha$,x)$^{121,123-126}$I reactions were studied. 

In addition, in the work of A. Saha~\citep{Saha2026}, $\alpha$-induced reaction cross-sections on Zn targets have been systematically investigated for the production of medically relevant isotopes using TALYS 1.96. Optimal combinations of nuclear level density and alpha optical model potentials are determined through mean-weighted deviation analysis. Complementarily, studies on $^{165}$Ho reactions~\citep{Tureci2025} incorporate TALYS and TENDL-2023 along with Deep Neural Network (DNN) models, demonstrating that machine learning techniques significantly enhance agreement with experimental results and improve cross-section predictions.

Although $\alpha$-induced reactions on medium-mass and light nuclei such as Zn~\citep{Saha2026}, Cu~\citep{Saha2025a}, and Ca~\citep{Saha2025b} have been previously investigated, extending these studies to heavier nuclei such as Ag provides a distinct and necessary contribution to the nuclear data community. The Ag system lies in a higher mass region (A$\sim$100), where nuclear level densities are significantly higher, and reaction mechanisms become more complex due to increased competition among multiple channels such as ($\alpha$,n), ($\alpha$,2n), and ($\alpha$,p).
This results in enhanced sensitivity of calculated cross sections to key nuclear reaction model ingredients, particularly $\alpha$-optical model potentials, nuclear level density, and pre-equilibrium contributions. Consequently, the Ag system serves as a rigorous testing ground for assessing the predictive performance and consistency of theoretical models across a heavier nuclear regime, in comparison to lighter systems such as Zn, Cu, and Ca.

\section{Methodology}

\subsection{Details of the experimental data taken from EXFOR}\label{sec2}

In the present work, experimental data reported by Misaelides {\em et al.}~\citep{Misaelides1980}, Guin {\em et al.}~\citep{Guin1992}, Wasilevsky {\em et al.}~\citep{Wasilevsky1985,Wasilevsky1986}, Fukushima {\em et al.}~\citep{Fukushima1963,Fukushima1965}, Tarkanyi {\em et al.}~\citep{Tarkanyi2015}, Yalcin {\em et al.}~\citep{Yalcin2015} and Singh {\em et al.}~\citep{Singh1987} taken from the EXFOR library have been used.

\subsection{Theoretical Background}\label{sec3}

In the present work, excitation functions of the $^{107}$Ag($\alpha$,xn)$^{108m,109g,110m}$In and $^{109}$Ag($\alpha$,xn)$^{110g,111g}$In reactions were estimated using TALYS 2.0 code. TALYS is a computer program which is used for the simulation of nuclear reactions involving neutrons, photons, protons, deuterons, tritons, $^3$He-and alpha-particles, in the 10$^{-11}$-1000 MeV energy range and for target nuclides of mass A in the 5$<$A$<$339 range~\citep{Koning2023}. TALYS 2.0 includes six Nuclear Level Density (NLD) models: three phenomenological models (Constant Temperature Fermi Gas Model~\citep{Gilbert1965,Dilg1973}, Back-shifted Fermi Gas Model~\citep{Gilbert1965,Dilg1973}, Generalized Superfluid Model~\citep{Ignatyuk1979,Ignatyuk1993}) and three microscopic models (Skyrme-Hartree-Fock-Bogoliubov~\citep{Goriely2001}, Gogny-Hartree-Fock-Bogoliubov~\citep{Goriely2008}, and Temperature-dependent Gogny-Hartree-Fock-Bogolyubov)~\citep{Hilaire2012}.

The nuclear level density, arising from both protons and neutrons, can be described as a function of the excitation energy $U$, the level density parameter $a$, the spin $J$, and the spin cut-off parameter $\sigma^2$~\citep{Bethe1937, Ericson1960}:

\begin{equation}
\rho(U, J, \Pi) = \frac{1}{2} \cdot \frac{2J + 1}{2\sqrt{2\pi}\sigma^3}
\exp\left(-\frac{(J + 1/2)^2}{2\sigma^2}\right)
\cdot \frac{\sqrt{\pi}}{12} \cdot \frac{\exp\left(2\sqrt{aU}\right)}{a^{1/4} U^{5/4}}
\end{equation}

In this expression, the spin cut-off parameter is defined as
\[
\sigma^2 = \frac{T I}{\hbar^2},
\]
where the nuclear temperature $T$ is given by
\[
T = \sqrt{\frac{U}{a}},
\]
and $I$ denotes the moment of inertia of the nucleus. 

The total level density is obtained by summing over all possible spin values, leading to

\begin{equation}
\rho^{\text{tot}}(U) = \frac{1}{12\sqrt{2}\sigma} \cdot \frac{\exp\left(2\sqrt{aU}\right)}{a^{1/4} U^{5/4}}.
\end{equation}

The different level density models are described below:- 

\begin{enumerate}

\item The Constant Temperature Fermi Gas Model~\citep{Gilbert1965,Dilg1973} (the default level density model) divides the excitation energy into two regions: a higher energy region where the Fermi gas model is applicable, and a lower energy region where the constant temperature law is useful. It is abbreviated as LDM1.

\item Back-Shifted Fermi Gas model~\citep{Gilbert1965,Dilg1973} (abbreviated as LDM2), which takes pairing energy and shell effects into account,  is used for predicting nuclear properties at lower excitation energies. 

\item The Generalized Superfluid Model~\citep{Ignatyuk1979,Ignatyuk1993} (abbreviated as LDM3) takes superconductive pairing correlations into account according to Bardeen-Cooper-Schrieffer theory. It combines superfluid effects at low excitation energies with Fermi-gas physics at high energies to describe the nuclear level density. 

\item Skyrme-Hartree-Fock-Bogoliubov~\citep{Goriely2001} (abbreviated as LDM4) method is based on Hartree-Fock calculations that provide level density parameters derived from nuclear structure physics. It generates a nuclear mass table from the drip line to the drip line by applying the Skyrme force. 

\item Gogny-Hartree-Fock-Bogoliubov model~\citep{Goriely2008} (abbreviated as LDM5) is another microscopic model based on Hartree-Fock calculations, often described as a combinatorial method developed by Hilaire and Goriely. This model determines the nuclear level densities for more than 8500 nuclei in a tabular format. 

\item Temperature-dependent Gogny-Hartree-Fock-Bogoliubov~\citep{Hilaire2012} (abbreviated as LDM6) is a more sophisticated model that uses the Gogny force to account for self-consistent pairing correlations and shell effects. It also accounts for how these properties change with excitation energy and temperature, enabling a better description of collective phenomena like rotations and vibrations. 

\end{enumerate}

Similarly, TALYS includes eight alpha optical model potentials ($\alpha$OMPs), which are used to describe alpha-induced reactions by modeling the complex interaction between an incoming alpha particle and a target nucleus. The generalized form of the alpha optical model potential is as follows:-

\begin{equation}
U(r)=V_c(r)+V(r)+iW(r)
\label{eqn1}
\end{equation}

where V$_c$(r) is a Coulomb potential, V(r) is the real, and W(r) is the imaginary part of the nuclear potential. There are eight different alpha optical model potentials available in the TALYS code, which are listed below:-

\begin{enumerate}

\item Watnabe folding approach with Koning-Delaroche nucleon potentials~\citep{Koning2003,Watanabe1958} (abbreviated as $\alpha$OMP1)-This approach is a theoretical framework used to derive the optical model potential for a composite projectile (like a deuteron, $^3$He, or 
$\alpha$-particle) by "folding" the individual nucleon-nucleus optical potentials of its constituent nucleons over the projectile's internal ground-state wave function
\item Alpha optical model potential of McFadden and Satchler~\citep{McFadden1966} (abbreviated as $\alpha$OMP2)-
The alpha optical model proposed by McFadden and Satchler describes the potential as consisting of real and imaginary components:

\begin{equation}
V(r) = -V \left(e^{\frac{r - R}{a}} + 1 \right)^{-1}
+ iW \left(e^{\frac{r - R}{a}} + 1 \right)^{-1}
\end{equation}

where
\[
R = r_0 A^{1/3}
\]
Here, $a$ represents the surface diffuseness parameter, and $r_0$ is the radius parameter. In this work, $r_c = r_0 A^{1/3} = 1.3\,\text{fm}$ is adopted for the Coulomb interaction. A=atomic mass number, V=depth of the real part of the nuclear potential, W=depth of the imaginary part of the potential, V and W are constants for a given projectile-target system and incident energy.
 
\item Double folding alpha optical model potential of Demetriou {\em et al.}~\citep{Demetriou2002} which comes with three possible options: two of them utilizes two imaginary potential tables (Table 1 and Table 2 of Demetriou {\em et al.}~\citep{Demetriou2002}) (abbreviated as $\alpha$OMP3, $\alpha$OMP4) and the Dispersive model of Demetriou {\em et al.}~\citep{Demetriou2002} (abbreviated as $\alpha$OMP5). $\alpha$OMP3 and $\alpha$OMP4 employ double-folding real potentials combined with different parameterizations of the imaginary potential based on tabulated systematics. $\alpha$OMP5 is a dispersive optical model that incorporates energy dependence through dispersion relations, ensuring a consistent link between real and imaginary components and improving performance at sub-barrier energies.

The dispersive alpha optical model proposed by Demetriou {\em et al.}~\citep{Demetriou2002} includes a damped surface term in the imaginary component of the potential. This term has an energy dependence and is expressed in a Woods--Saxon form:

\begin{equation}
W(r, E) = W_{0,v} \, F_s(E) \, f_v(r)
\end{equation}

where the subscript $v$ denotes the contribution of the volume. The radial form factor is given by

\begin{equation}
f_v(r) = \frac{1}{1 + \exp\left(\frac{r - r_v A^{1/3}}{a_v}\right) }
\end{equation}

This is a similar radial dependence as above.
The energy-dependent factor is defined as

\begin{equation}
F_s(E) = F(E)\,\exp\left[-C_s \, |E - E_f|\right]
\end{equation}

where $E_f$ is the Fermi energy. The global dispersive model of Demetriou uses a Coulomb radius $r_c = 1.25\,\text{fm}$. The function $F(E)$ represents the Fermi-type parametrization, and $C_s = 0.01951 - 0.00049E$ is an energy-dependent parameter obtained by fitting experimental data.

\item Alpha optical model potential of Avrigeanu {\em et al.}~\citep{Avrigeanu2014} (the default one) (abbreviated as $\alpha$OMP6), Alpha optical model potentials of Nolte {\em et al.}~\citep{Nolte1987} (abbreviated as $\alpha$OMP7) and Alpha optical model potential of Avrigeanu {\em et al.}~\citep{Avrigeanu1994} (abbreviated as $\alpha$OMP8). $\alpha$OMP6 is a comprehensive phenomenological potential with explicit energy and mass dependence, optimized using extensive experimental data on alpha-induced reactions. It provides reliable predictions over a broad energy range, particularly for medium-mass nuclei~\citep{Avrigeanu2014}. $\alpha$OMP7 is developed primarily for higher-energy alpha scattering; this potential is characterized by a stronger absorptive imaginary component, making it suitable for reactions where surface absorption dominates. $\alpha$OMP8 is an extension of the Avrigeanu potential with refined parameter sets for improved agreement with specific experimental datasets, particularly in near-barrier and low-energy regions.

\end{enumerate}

Similarly, four different types of pre-equilibrium models are available in the TALYS 2.0 code, which are as follows:-

\begin{enumerate}

\item PE1: exciton model~\citep{Kalbach1986} where the analytical transition rates are used along with the energy-dependent matrix element

\item PE2: exciton model in which numerical transition rates are used with energy-dependent matrix elements. This is the default model for describing reactions where incoming particles interact with a few nucleons before forming a compound nucleus. This model treats particles and holes of both neutrons and protons through successive particle-hole configurations for the determination of pre-equilibrium emission rate of particles.

\item PE3: exciton model in which numerical transition rates are used with the optical model for collision probability

\item PE4: multi-step direct or compound model 

The exciton model (i.e., PE1, PE2 and PE3) inherently accounts for contributions from pre-equilibrium reactions involving deuterons, tritons, helium-3, and alpha particles. However, it is well known ~\citep{Blideanu2004} that nuclear reactions with differing particle numbers in the entrance and exit channels-such as stripping, pick-up, break-up, and knock-out processes-are not described within the exciton model framework.

To address these mechanisms, Kalbach~\citep{Kalbach2005} introduced a phenomenological contribution, which has been implemented in TALYS. Accordingly, the total pre-equilibrium cross section is expressed as the sum of contributions from the exciton model (EM), nucleon transfer (NT), and knock-out (KO) processes:
\begin{equation}
\frac{d\sigma_k^{\mathrm{PE}}}{dE_k}
=
\frac{d\sigma_k^{\mathrm{EM}}}{dE_k}
+
\frac{d\sigma_k^{\mathrm{NT}}}{dE_k}
+
\frac{d\sigma_k^{\mathrm{KO}}}{dE_k}.
\end{equation} 
\end{enumerate}

\subsection{Computational Details of TALYS Calculations}\label{uses}
The nuclear reaction calculations were performed using the TALYS-2.0 code in a predictive mode without any adjustment of model parameters to the experimental data. The excitation functions for the studied reactions were generated by systematically employing different combinations of alpha optical model potentials ($\alpha$OMP1--$\alpha$OMP8), level density models (LDM1--LDM6), and pre-equilibrium models (PE1--PE4) available within TALYS.
All calculations were carried out over the same incident energy range as the experimental datasets to ensure consistent comparison. Therefore, no fixed or uniform energy step size was employed in the calculations. The energy range for each reaction is determined by the minimum and maximum energies covered by the respective experimental data sets. The resulting theoretical cross sections were compared with experimental data, and the agreement was quantified using the reduced chi-square metric ($\chi^2_\nu$) defined using the following formula:- 

\begin{equation}
\chi^2_\nu = \frac{1}{\nu} \sum_{i=1}^{N}
\frac{\left(\sigma_i^{\mathrm{expt}} - \sigma_i^{\mathrm{sim}}\right)^2}
{\left(\Delta \sigma_i^{\mathrm{expt}}\right)^2}.
\end{equation}

where $\sigma_i^{expt}$=experimental cross section for i$^{th}$ data point, $\sigma_i^{sim}$=simulated cross section for i$^{th}$ data point, N= no. of experimental data points. Here, the number of degrees of freedom is given by $\nu = N - 1$. In the present work, the quantity $\chi^2_\nu$ is employed as a normalized goodness-of-fit metric for comparative ranking of the 192 TALYS model combinations. Since no model parameters are adjusted to the experimental data, the choice between N and N-1 introduces only a constant scaling factor within a given reaction channel and does not affect model rankings, sensitivity indices, or the resulting conclusions.

This systematic approach enables the evaluation of the sensitivity of the calculated cross sections to different nuclear model inputs and facilitates the identification of the most suitable model combinations. In the present work, TALYS-based theoretical calculation of cross section has been carried out using each of the 192 possible combinations out of the six level density models, eight alpha optical model potentials, and four pre-equilibrium models as discussed above. The calculated theoretical cross-sections are then compared with the experimental data taken from the EXFOR library. The goodness of fit between the theoretical and experimental data is quantified using the $\chi^2_\nu$ analysis. All combinations of the level density model, alpha optical model potential, and pre-equilibrium model that yield the $\chi^2_\nu$ value within 5\% of the minimum value are considered as statistically equivalent better-peforming combinations within the adopted TALYS 2.0 framework rather than a single point optimum for describing the experimental data.

However, in the present work, we encountered a practical limitation associated with the EXFOR database: among the EXFOR datasets considered in the present work, only the measurements of S. Fukushima {\em et al.}~\citep{Fukushima1963,Fukushima1965} contain entries with zero reported uncertainties. The direct use of such values in the standard expression leads to divergent chi-square values that are not physically meaningful and prevent a consistent comparison across datasets. To address this issue, for data points with zero uncertainties in EXFOR, a minimum relative uncertainty of 5\% has been assigned to avoid divergence in the chi-square calculation, consistent with standard practices in nuclear data evaluation.

Since the uncertainty floor directly affects the weighting of experimental points lacking reported uncertainties, an additional sensitivity test was performed by increasing the minimum relative uncertainty from 5\% to 10\%. The corresponding results are summarized in Table~\ref{tab:uncertainty_floor}. Although the numerical sensitivity percentages change slightly because of the reduced statistical weight of the affected data, the dominant sensitivity drivers and the qualitative interpretation remain unchanged for all three reaction channels. The conclusions regarding the governing model ingredients are, therefore, robust with respect to the adopted uncertainty floor.  

\section{Results and Discussion}\label{sec4}

The threshold energies (E$_{th}$) for the reactions studied, together with the corresponding Q-values, are listed in Table~\ref{tab6}. It is well known that reaction cross sections exhibit strong sensitivity to nuclear model parameters in the vicinity of the threshold energy, where the opening of reaction channels leads to a rapid increase in the cross section. In this energy region, small variations in the level density, optical model potential, and pre-equilibrium contributions can lead to significant differences in the calculated cross sections, thereby influencing the goodness-of-fit between the model calculations and experimental data. At higher incident energies, where multiple channels are fully open, the dependence on model parameters becomes comparatively smoother, leading to a more stable optimization behavior. For each reaction channel, the number of experimental data sets, the number of data points per set (N), the year of measurement, the energy range covered, and the reported uncertainties are listed in Table~\ref{tab7}.

\begin{table}[h!]
\centering
\caption{Threshold energies and Q-values for the studied reactions}
\begin{tabular}{|ccc|}
\hline
\textbf{Reaction} & \textbf{Q-value (MeV)} & \textbf{Threshold Energy $E_{\text{th}}$ (MeV)} \\
\hline
$^{107}$Ag($\alpha$,3n)$^{108\text{m}}$In & -26.1 & 27.1 \\
$^{107}$Ag($\alpha$,2n)$^{109g}$In & -15.6 & 16.22 \\
$^{109}$Ag($\alpha$,3n)$^{110\text{g}}$In & -24 & 24.9 \\
$^{107}$Ag($\alpha$,n)$^{110\text{m}}$In & -7.65 & 7.93 \\
$^{109}$Ag($\alpha$,2n)$^{111\text{g}}$In & -14 & 14.6 \\
\hline
\end{tabular}\label{tab6}
\end{table}

\begin{table*}[h!]
\centering
\caption{Compilation of experimental datasets used for the studied reactions, including references, number of data points, energy ranges, and reported uncertainties.}
\renewcommand{\arraystretch}{1.2}
\begin{tabular}{c c c c c c}
\toprule
\textbf{Reaction} & \textbf{Dataset} & \textbf{N} & \textbf{N (total)} & \textbf{Energy Range (MeV)} & \textbf{Uncertainty (\%)} \\
\midrule

\multirow{3}{*}{$^{107}$Ag($\alpha$,3n)$^{108\text{m}}$In}
& \citep{Misaelides1980}   & 4  & \multirow{3}{*}{23} & \multirow{3}{*}{30.3--55.9} & $\sim$16--20 \\
& \citep{Wasilevsky1985}   & 10 &  &  & $\sim$14--45 \\
& \citep{Guin1992}         & 9  &  &  & $\sim$7--14 \\
\midrule

\multirow{1}{*}{$^{107}$Ag($\alpha$,2n)$^{109\text{g}}$In}
& \citep{Fukushima1963}   & 23 & \multirow{1}{*}{23} & \multirow{1}{*}{16.6--39.4} & $\sim$5--36 \\
\midrule

\multirow{4}{*}{$^{109}$Ag($\alpha$,3n)$^{110\text{g}}$In}
& \citep{Misaelides1980}   & 6  & \multirow{4}{*}{37} & \multirow{4}{*}{26.5--58.4} & $\sim$10--22 \\
& \citep{Fukushima1965}   & 13 &  &  & $\sim$5--29 \\
& \citep{Singh1987}        & 7  &  &  & $\sim$13 \\
& \citep{Guin1992}         & 11 &  &  & $\sim$7--9 \\
\midrule

\multirow{4}{*}{$^{107}$Ag($\alpha$,n)$^{110\text{m}}$In}
& \citep{Fukushima1963}    & 20 & \multirow{4}{*}{47} & \multirow{4}{*}{8.9--41.6} & $\sim$5--27 \\
& \citep{Misaelides1980}   & 6  &  &  & 13--32 \\
& \citep{Tarkanyi2015}     & 14 &  &  & 4--6 \\
& \citep{Yalcin2015}       & 11 &  &  & $\sim$7--18 \\
\midrule

\multirow{2}{*}{$^{109}$Ag($\alpha$,2n)$^{111\text{g}}$In}
& \citep{Wasilevsky1986}   & 13 & \multirow{2}{*}{22} & \multirow{2}{*}{20--54.2} & $\sim$14--50 \\
& \citep{Guin1992}         & 9  &  &  & $\sim$7--13 \\
\bottomrule

\end{tabular}\label{tab7}
\end{table*}

The values of the reduced $\chi^2$ estimated for the $^{107}$Ag($\alpha$,3n)$^{108m}$In, $^{107}$Ag($\alpha$,2n)$^{109g}$In, $^{107}$Ag($\alpha$,n)$^{110m}$In, $^{109}$Ag($\alpha$,3n)$^{110g}$In, and $^{109}$Ag($\alpha$,2n)$^{111g}$In reactions with respect to the experimental data from EXFOR are presented in Supplementary Tables~\ref{tab1}, \ref{tab2}, \ref{tab3}, \ref{tab4}, and \ref{tab5}, respectively, in the Appendix. Figures~\ref{fig1}--\ref{fig5} show comparisons between the results obtained using the better-performing model combination with minimum $\chi^2_\nu$, the default TALYS calculation, and a model with a higher $\chi^2_\nu$ value, along with the EXFOR experimental data and TENDL-2023 evaluated data. To provide a qualitative assessment of the model dependence of the calculated excitation functions, model-dispersion bands have been constructed from the complete ensemble of 192 TALYS calculations performed for each reaction channel. At every incident alpha energy, the minimum and maximum calculated cross sections among all parameter combinations were determined, and the region enclosed by these values is displayed as a shaded band in the excitation-function plots. These bands represent the spread arising from the investigated variations of the TALYS statistical-model parameters and are intended as an indicator of model sensitivity rather than as a rigorous statistical uncertainty interval.

Although the model-dispersion band encompasses a wide range of theoretical predictions arising from different combinations of optical model potentials, level density models and pre-equilibrium models, none of the 192 TALYS-2.0 model combinations quantitatively reproduces the experimental excitation functions over the investigated energy range. This inability to achieve a satisfactory quantitative agreement suggests that additional nuclear-reaction physics or improved parameterizations may be required for the $A\approx100$ mass region.

\begin{table*}[htbp]
\centering
\caption{Comparison of the best-fit TALYS parameter combination and the TALYS parameter combination whose reduced $\chi^2$ with respect to the experimental data most closely matches that of the TENDL-2023 evaluation. The corresponding reduced $\chi^2$ values for the best TALYS calculation and TENDL-2023 are also presented.}
\label{tab:tendl_compare}

\renewcommand{\arraystretch}{1.3}

\begin{tabular}{|p{2.3cm}|p{3.3cm}|c|c|p{3.8cm}|}
\hline
\textbf{Reaction} &
\textbf{Best TALYS combination} &
\textbf{$\chi^2_\nu$ (Best TALYS)} &
\textbf{$\chi^2_\nu$ (TENDL-2023)} &
\textbf{closest TALYS combination } \\
\hline

$^{107}$Ag($\alpha$,3n)$^{108m}$In &
LDM6+$\alpha$OMP1+PE3 &
48 &
120 &
LDM2+$\alpha$OMP4+PE2,\\
&&&&LDM2+$\alpha$OMP5+PE2,\\
&&&&LDM1+$\alpha$OMP2+PE3 \\
\hline

$^{107}$Ag($\alpha$,2n)$^{109g}$In &
LDM6+$\alpha$OMP6+PE3 &
25 &
93 &
LDM1+$\alpha$OMP1+PE4,\\
&&&&LDM1+$\alpha$OMP4+PE4,\\
&&&&LDM1+$\alpha$OMP5+PE4,\\
&&&&LDM2+$\alpha$OMP8+PE4 \\
\hline

$^{107}$Ag($\alpha$,n)$^{110m}$In &
LDM3+$\alpha$OMP2+PE2 &
60 &
130 &
LDM2+$\alpha$OMP1+PE3,\\
&&&&LDM2+$\alpha$OMP5+PE3 \\
\hline

$^{109}$Ag($\alpha$,3n)$^{110g}$In &
LDM1+$\alpha$OMP7+PE3 &
34 &
76 &
LDM5+$\alpha$OMP7+PE3 \\
\hline

$^{109}$Ag($\alpha$,2n)$^{111g}$In &
LDM6+$\alpha$OMP8+PE4 &
21 &
183 &
LDM5+$\alpha$OMP2+PE1 \\
\hline

\end{tabular}
\end{table*}

\begin{table}[htbp]
\centering
\caption{Summary of better-performing TALYS model combinations for $\alpha$-induced reactions on Ag isotopes. The minimum $\chi^2_\nu$ ranges (within 5\% tolerance) and corresponding preferred level density models, alpha optical model potentials, and pre-equilibrium models are shown. The clustering of combinations reflects the dependence of model preference on reaction channel and excitation energy.}
\vspace{0.5cm}

\begin{tabular}{|cccccc|}
\hline
\textbf{Reaction} & \textbf{$\chi^2_\nu$ (min.) Range} & \textbf{Preferred LDM} & \textbf{OMP Range} & \textbf{PE Models} & \textbf{No. of Comb.} \\
\hline

$^{107}$Ag($\alpha$,3n)$^{108m}$In & 48--50 & LDM6 & OMP1--OMP8 & PE1--PE3 & 24 \\
\hline

$^{107}$Ag($\alpha$,2n)$^{109g}$In & 25--26 & LDM5--LDM6 & OMP6 & PE3 & 2 \\
\hline

$^{107}$Ag($\alpha$,n)$^{110m}$In & 60--63 & LDM3 & OMP1--OMP5 & PE1--PE3 & 15 \\
\hline

$^{109}$Ag($\alpha$,3n)$^{110g}$In & 34--124 & LDM1 & OMP7--OMP8 & PE1--PE3 & 6 \\
\hline

$^{109}$Ag($\alpha$,2n)$^{111g}$In & 21--22 & LDM5--LDM6 & OMP7--OMP8 & PE4 & 4 \\
\hline

\end{tabular}\label{tab8}
\end{table}

\begin{figure}[H]
\centering
\includegraphics[width=0.6\textwidth]{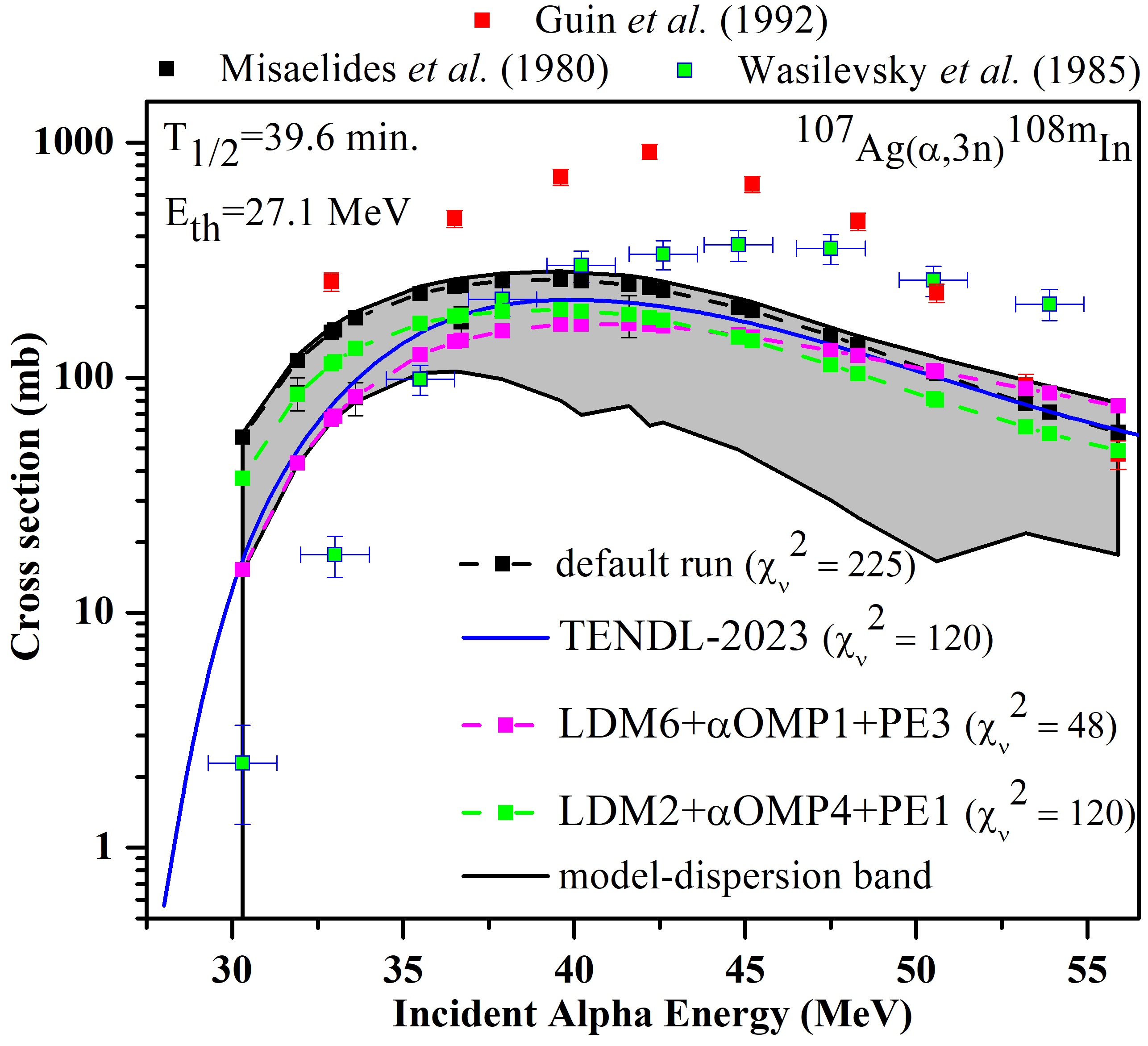}
\caption{(color online) Comparison of calculated cross sections for the $^{107}$Ag($\alpha$,3n)$^{108m}$In reaction using different combinations of nuclear level density models, alpha optical model potentials, and pre-equilibrium models. Results corresponding to the minimum $\chi^2_\nu$, the default run, and a higher $\chi^2_\nu$ combination are shown alongside experimental data reported by Wasilevsky et al.~\citep{Wasilevsky1985}, Misaelides et al.~\citep{Misaelides1980}, and Guin et al.~\citep{Guin1992}, taken from the EXFOR database~\citep{Otuka2014}, as well as evaluated data from TENDL-2023~\citep{Koning2019}. The comparison highlights the sensitivity of the cross-section predictions to the choice of input model combinations. The gray shaded region represents the model-dispersion band constructed from the complete set of 192 TALYS calculations. At each incident alpha energy, the band is bounded by the minimum and maximum calculated cross sections and provides a qualitative measure of the model dependence arising from variations in the adopted statistical-model parameters.
}\label{fig1}
\end{figure}

\begin{figure}[H]
\centering
\includegraphics[width=0.6\textwidth]{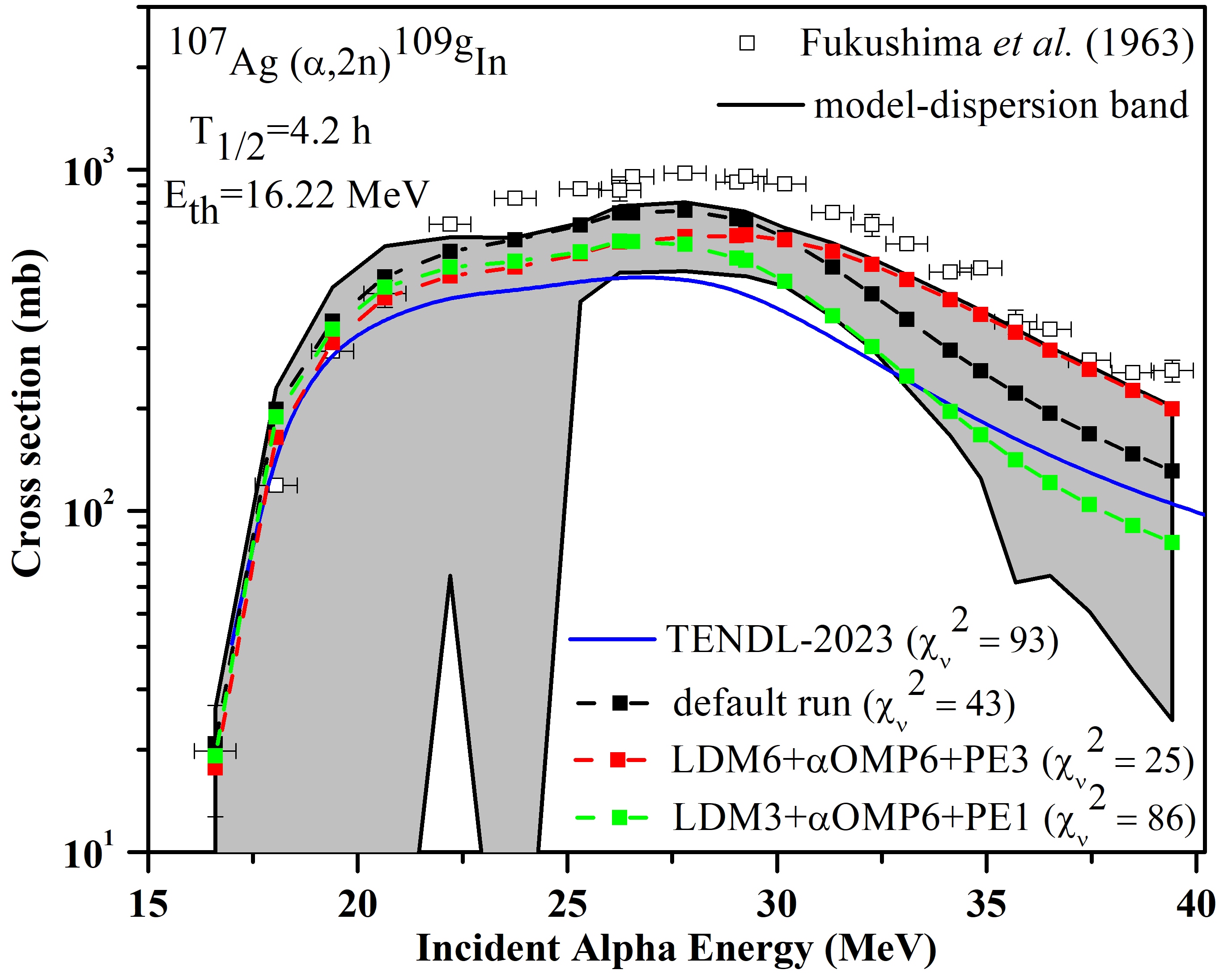}
\caption{(color online) Comparison of calculated cross sections for the $^{107}$Ag($\alpha$,2n)$^{109g}$In reaction using different combinations of nuclear level density models, alpha optical model potentials, and pre-equilibrium models. Results corresponding to the minimum $\chi^2_\nu$, the default run, and a higher $\chi^2_\nu$ combination are shown alongside experimental data reported by Fukushima {\em et al.}~\citep{Fukushima1963}, taken from the EXFOR database~\citep{Otuka2014}, as well as evaluated data from TENDL-2023~\citep{Koning2019}. It is to be noted that this single dataset forms the entire experimental basis for the optimization.
}\label{fig2}
\end{figure}

\begin{figure}[H]
\centering
\includegraphics[width=0.6\textwidth]{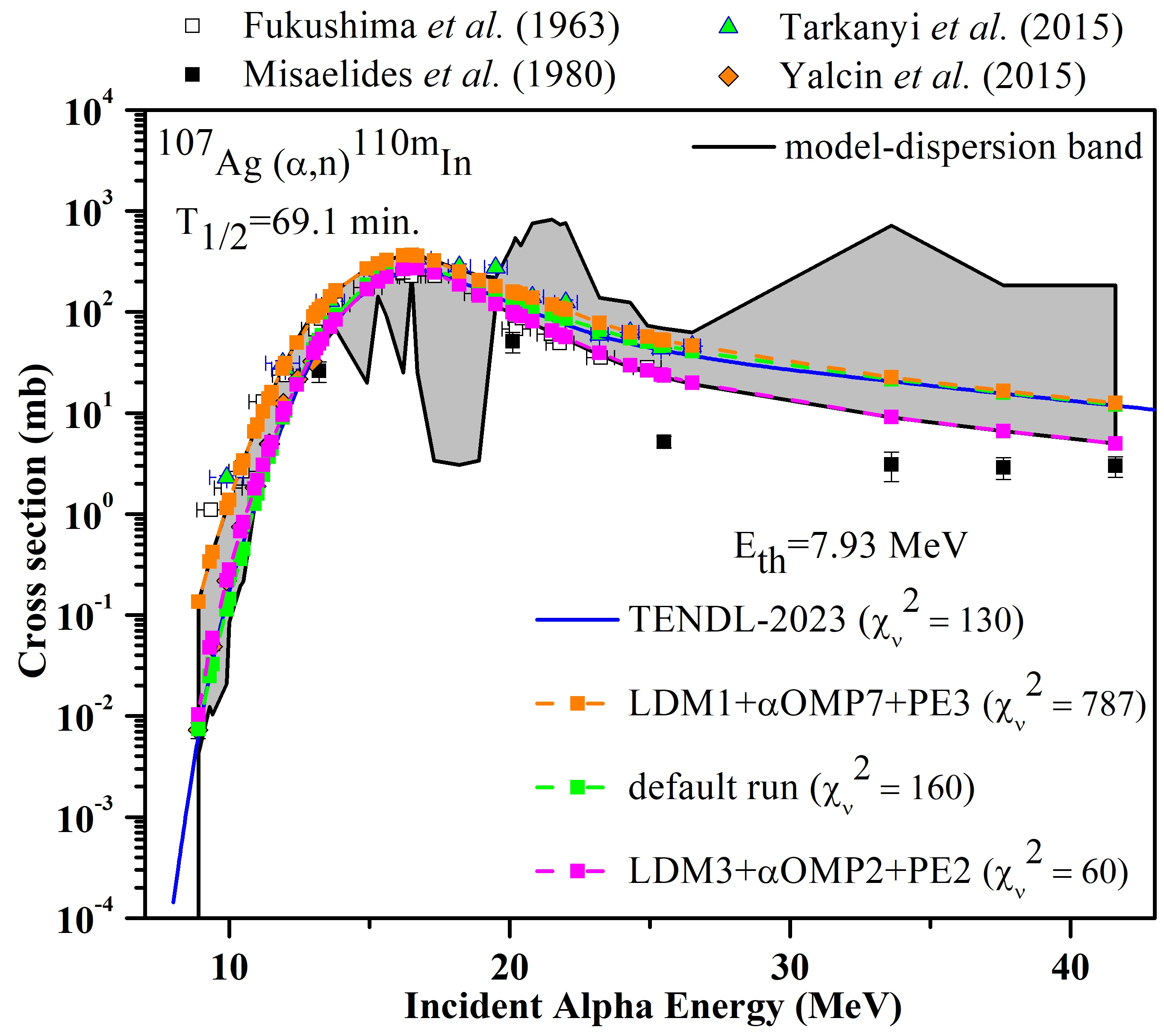}
\caption{(color online) Same as Fig.~\ref{fig1} and Fig.~\ref{fig2}, but for the $^{107}$Ag($\alpha$,n)$^{110m}$In reaction. Experimental data reported by Fukushima {\em et al.}~\citep{Fukushima1963}, Misaelides {\em et al.}~\citep{Misaelides1980}, Tarkanyi {\em et al.}~\citep{Tarkanyi2015}, and Yalcin {\em et al.}~\citep{Yalcin2015} are taken from the EXFOR database~\citep{Otuka2014}.}\label{fig3}
\end{figure}

\begin{figure}[H]
\centering
\includegraphics[width=0.6\textwidth]{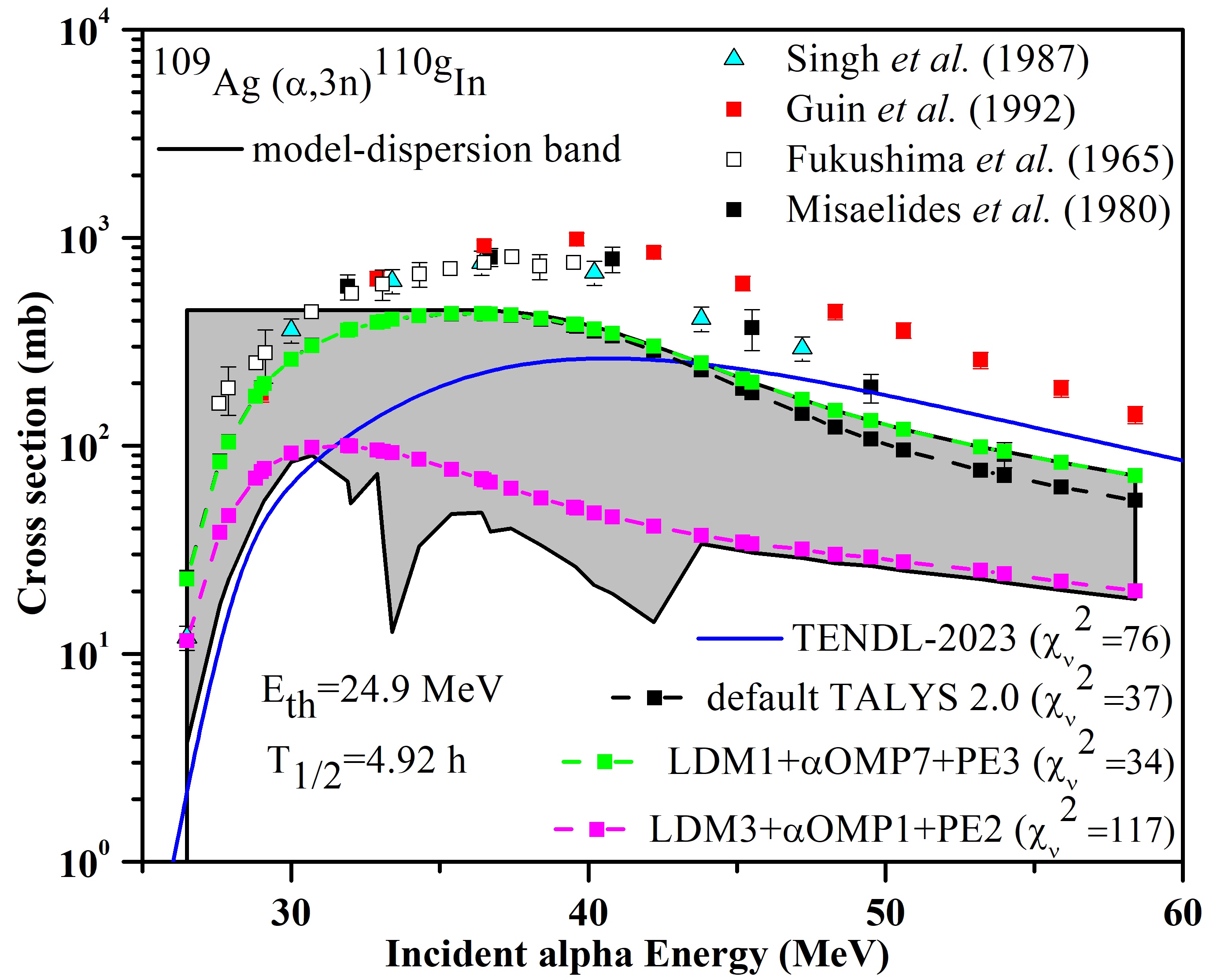}
\caption{(color online) Same as Fig.~\ref{fig1} and Fig.~\ref{fig2}, but for the $^{109}$Ag($\alpha$,3n)$^{110g}$In reaction. Experimental data reported by Fukushima {\em et al.}~\citep{Fukushima1965}, Misaelides {\em et al.}~\citep{Misaelides1980}, Singh {\em et al.}~\citep{Singh1987}, and Guin {\em et al.}~\citep{Guin1992} are taken from the EXFOR database~\citep{Otuka2014}.
 }\label{fig4}
\end{figure}

\begin{figure}[H]
\centering
\includegraphics[width=0.6\textwidth]{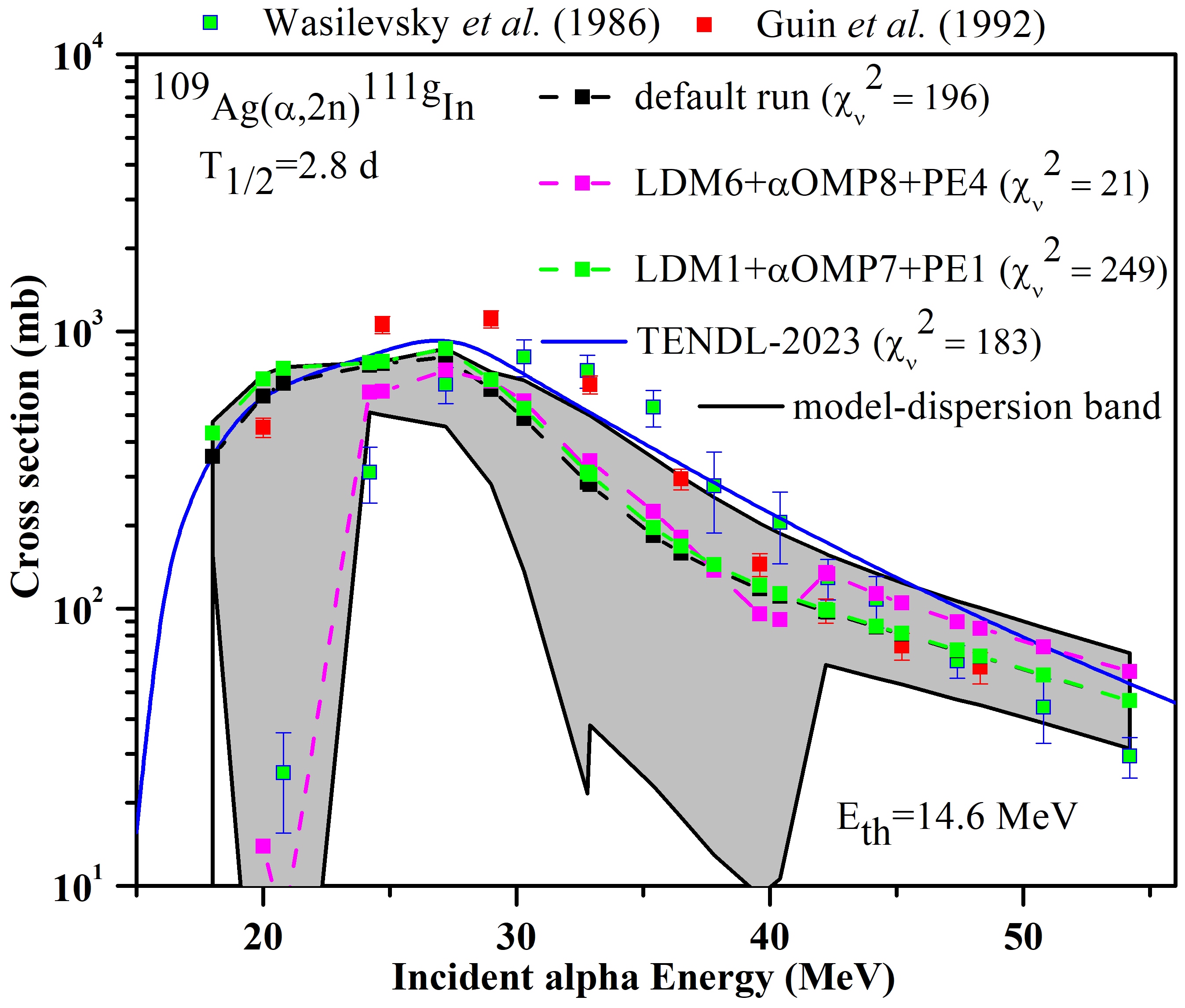}
\caption{(color online) Same as Fig.~\ref{fig1} and Fig.~\ref{fig2}, but for the $^{109}$Ag($\alpha$,2n)$^{111g}$In reaction. Experimental data reported by Wasilevsky {\em et al.}~\citep{Wasilevsky1986} and Guin {\em et al.}~\citep{Guin1992} are taken from EXFOR library~\citep{Otuka2014}.}\label{fig5}
\end{figure}

To facilitate comparison with the TALYS-based TENDL-2023 evaluation, Table~\ref{tab:tendl_compare} summarizes, for each reaction channel, the best-fitting TALYS model combination with respect to the experimental data together with the model combination whose reduced $\chi^2$ value (relative to the experimental data) is numerically closest to that obtained for the TENDL-2023 evaluation. The corresponding reduced $\chi^2$ values for the best TALYS calculation and the TENDL-2023 evaluation are also listed. This comparison provides an indication of how closely the optimized TALYS calculations reproduce the agreement with the experimental data achieved by the TENDL-2023 evaluation.

A systematic analysis of the calculated $\chi^2_\nu$ values within a 5-\% tolerance band, as shown in Table~\ref{tab8}, reveals well-defined patterns in the better-performing combinations of level density models, alpha optical model potentials, and pre-equilibrium formalisms across the investigated reaction channels. For multi-neutron emission channels such as ($\alpha$,2n) and ($\alpha$,3n), the solutions with comparatively improved agreement are predominantly clustered around higher level density prescriptions (i.e., LDM5-LDM6) coupled with a broad range of global OMP parameterizations (OMP1-OMP8) and PE1-PE3 models. This clustering indicates that, at higher excitation energies, the calculated cross sections are primarily governed by the statistical properties of the compound nucleus, with reduced sensitivity to the fine details of the optical potential. In contrast, the ($\alpha$,n) channel consistently favors formulations with lower excitation energy level density, such as LDM3, with comparatively tighter constraints on acceptable OMP-PE combinations, reflecting enhanced sensitivity to nuclear structure effects near the reaction threshold. Furthermore, the limited appearance of PE4 among the acceptable solutions suggests that models incorporating strong multi-step direct components are not well suited for the present $\alpha$-induced reactions, where compound nucleus formation remains the dominant mechanism. The observed grouping of better-performing solutions across all five reactions thus provides strong evidence that the preference of the model is primarily dictated by the reaction channel and the corresponding excitation energy regime, rather than by a unique or isolated choice of model parameters.
	
\section{Robustness analysis of preferred model selections}

To quantify parameter sensitivity, the marginal variation of $\chi^2$ along each model axis—level density model, $\alpha$-optical model potential, and pre-equilibrium—was evaluated. For a given parameter, $\Delta \chi^2$ is defined as the difference between the maximum and minimum $\chi^2$ values obtained by varying that parameter while averaging (arithmetic mean) over all combinations of the remaining parameters. 

The relative sensitivity of each model component was then expressed in terms of normalized contributions:
\begin{equation}
S_i = \frac{\Delta \chi_i^2}{\sum_j \Delta \chi_j^2}
\end{equation}
where $i$ denotes the model axis. The percentage contribution is given by $S_i \times 100$. This approach enables the identification of the dominant physics inputs governing each reaction channel.

\begin{table*}[h!]
\centering
\caption{Relative contributions of different nuclear reaction model ingredients-level density models, $\alpha$-optical model potentials, and pre-equilibrium mechanisms-quantified in terms of normalized sensitivity $S_i$ (in \%) derived from the marginal variation $\Delta \chi^2$ for the investigated reactions. Here, $\Delta \chi^2$ represents the difference between the maximum and minimum $\chi^2$ values obtained by varying a given model parameter while averaging (arithmetic mean) over all combinations of the remaining parameters.}
\vspace{1cm}
\renewcommand{\arraystretch}{1.3}
\begin{tabular}{c c c c p{5.5cm}}
\toprule
Reaction & $S_{\mathrm{LDM}}$ (\%) & $S_{\alpha\mathrm{OMP}}$ (\%) & $S_{\mathrm{PE}}$ (\%) & Interpretation \\
\midrule
$^{107}$Ag($\alpha$,3n)$^{108\text{m}}$In   & 87.5 & 7.6 & 4.9 & primarily governed by LDM \\

$^{107}$Ag($\alpha$,2n)$^{109\text{g}}$In   & 42.7 & 11.3 & 46.0 & LDM and PE are comparable \\

$^{109}$Ag($\alpha$,3n)$^{110\text{g}}$In   & 85 & 4 & 11 & primarily governed by LDM \\

$^{107}$Ag($\alpha$,n)$^{110\text{m}}$In   & 5.1 & 31.7 & 63.2 & \makecell[l]{dominated by PE, $\alpha$OMP playing secondary role, \\ LDM contributing minimally} \\

$^{109}$Ag($\alpha$,2n)$^{111\text{g}}$In   & 14.9 & 21.9 & 63.2 & \makecell[l]{dominated by PE, $\alpha$OMP playing secondary role, \\ LDM contributing minimally} \\
\bottomrule
\end{tabular}\label{tab9}
\end{table*}

\section{Physics Interpretation and Cross-System Analysis}

The data presented in Table~\ref{tab9} quantify the sensitivity of the calculated cross sections, as reflected through variations in $\chi^2$, to changes in three principal nuclear reaction model ingredients: the level density models, the $\alpha$-optical model potential, and the pre-equilibrium mechanism. The reported percentages correspond to the normalized sensitivities $S_i$ derived from $\Delta \chi^2$.

For the $^{107}$Ag($\alpha$,3n)$^{108m}$In and
$^{109}$Ag($\alpha$,3n)$^{110g}$In reaction channels, the sensitivity is dominated by the LDM ($\sim$85--88\%), indicating that these multi-neutron emission reactions are governed primarily by compound-nucleus formation and its subsequent statistical decay, with comparatively minor contributions from the $\alpha$OMP and PE models.

The $^{107}$Ag($\alpha$,2n)$^{109g}$In reaction exhibits nearly equal sensitivities to the LDM and PE model ($\sim$43\% and $\sim$46\%, respectively), indicating that both compound-nucleus decay and pre-equilibrium emission contribute significantly to the calculated cross sections.

In contrast, the $^{107}$Ag($\alpha$,n)$^{110m}$In and
$^{109}$Ag($\alpha$,2n)$^{111g}$In reactions are predominantly influenced by the PE mechanism ($\sim$63\%), although the role of the $\alpha$OMP differs between the two channels. For the $^{107}$Ag($\alpha$,n)$^{110m}$In reaction, the $\alpha$OMP contributes substantially ($\sim$32\%), whereas the LDM contribution is minimal ($\sim$5\%). In the $^{109}$Ag($\alpha$,2n)$^{111g}$In reaction, the $\alpha$OMP plays a secondary role ($\sim$22\%), while the LDM contribution remains comparatively small ($\sim$15\%). These observations indicate an enhanced importance of direct reaction dynamics in these channels compared with the predominantly compound-nucleus character of the multi-neutron emission reactions.

The PE4 pre-equilibrium option exhibits pronounced channel-dependent behaviour across the investigated reactions. While PE4 yields the lowest reduced $\chi^2$ value for the $^{109}$Ag($\alpha$,2n)$^{111g}$In channel, it performs poorly for several other reaction channels, particularly $^{107}$Ag($\alpha$,n)$^{110m}$In, where substantially larger $\chi^2$ values are obtained. This inconsistency suggests that the apparent preference for PE4 in a single reaction channel should be interpreted with caution. Within the present TALYS~2.0 parameter space, the result indicates a localized improvement in agreement with the experimental data rather than definitive evidence for the physical superiority of the PE4 option. Additional investigations incorporating model uncertainties and independent validation would be required before attributing physical significance to the observed preference for PE4.

A comparative analysis of the result of the current $\alpha$-induced reaction on Ag isotopes with previous studies on the Mo target~\citep{Canbula2024} and the Zn target~\citep{Saha2026} shows a consistent trend in model performance. In all these cases, satisfactory agreement with experimental cross sections is achieved only through the combined use of appropriate alpha optical model potentials, level density models, and pre-equilibrium formalisms within TALYS. In particular, multi-particle emission channels like ($\alpha$,p3n), ($\alpha$,2p3n), ($\alpha$,2n) and ($\alpha$,3n) are better reproduced using microscopic or semi-microscopic level density prescriptions such as LDM6 in combination with suitable $\alpha$OMP and inclusion of PE effects (refer Table~\ref{tab10}), while the single particle channels like ($\alpha$,n), ($\alpha$,p) etc. exhibits comparatively higher sensitivity to the choice of models. Despite variations in the better-performing model combinations for the Zn, Mo, and Ag systems, the overall behavior remains consistent, indicating that the observed model preference is primarily governed by the reaction mechanism rather than the target mass.\\
\begin{table}[h]
\centering
\caption{Optimal combinations of LDM and $\alpha$OMP for different reaction channels based on earlier works~\citep{Canbula2024,Saha2026}}
\vspace{0.5cm}
\begin{tabular}{|ccc|}
\hline
\textbf{Reactions} & \textbf{Optimal LDM} & \textbf{Optimal $\alpha$OMP} \\
\hline
$^{64}$Zn($\alpha$,pn)$^{66}$Ga & LDM3 & $\alpha$OMP7 \\
$^{64}$Zn($\alpha$,p)$^{67}$Ga & LDM6 & $\alpha$OMP5 \\

$^{66}$Zn($\alpha$,p2n)$^{67}$Ga & LDM6 & $\alpha$OMP7 \\

$^{66}$Zn($\alpha$,2n)$^{68}$Ge & LDM6 & $\alpha$OMP3 \\
$^{66}$Zn($\alpha$,n)$^{69}$Ge & LDM2 & $\alpha$OMP3 \\

$^{66}$Zn($\alpha$,2p3n)$^{65}$Zn & LDM6 & $\alpha$OMP3 \\

$^{67}$Zn($\alpha$,p3n)$^{67}$Ga & LDM6 & $\alpha$OMP7 \\
$^{67}$Zn($\alpha$,3n)$^{68}$Ge & LDM6 & $\alpha$OMP3 \\
$^{68}$Zn($\alpha$,3n)$^{69}$Ge & LDM4 & $\alpha$OMP5 \\
$^{92}$Mo($\alpha$,p5n)$^{90}$Nb & LDM4 & $\alpha$OMP8 \\
$^{94}$Mo($\alpha$,2n)$^{96}$Tc & LDM6 & $\alpha$OMP5 \\
$^{95}$Mo($\alpha$,3n)$^{96}$Tc & LDM6 & $\alpha$OMP6 \\
\hline
\end{tabular}

\label{tab10}
\end{table}

\noindent\textbf{Robustness of the sensitivity analysis to the assumed uncertainty floor.}

From the datasets of~\cite{Fukushima1963,Fukushima1965}, 17 of 23, 8 of 36, and 17 of 47 EXFOR data points lacked reported experimental uncertainties for the $^{107}$Ag($\alpha$,2n)$^{109g}$In, $^{109}$Ag($\alpha$,3n)$^{110g}$In, and $^{107}$Ag($\alpha$,n)$^{110m}$In reactions, respectively. These data points were assigned a minimum relative uncertainty of 5\% because no experimental uncertainties were available. To examine the sensitivity of the results to this assumption, the analysis was repeated using a 10\% uncertainty floor. As summarized in Table~\ref{tab:uncertainty_floor}, increasing the assumed minimum uncertainty produces only modest changes in the normalized sensitivity values and does not alter the qualitative interpretation for any of the three reaction channels. In particular, the $^{107}$Ag($\alpha$,2n)$^{109g}$In reaction continues to exhibit comparable sensitivities to the level-density model and the pre-equilibrium mechanism, while the $^{109}$Ag($\alpha$,3n)$^{110g}$In reaction remains primarily governed by the level-density model. Likewise, the $^{107}$Ag($\alpha$,n)$^{110m}$In reaction continues to be dominated by the pre-equilibrium mechanism, with the $\alpha$-optical model potential providing a secondary contribution and the level-density model contributing only minimally. Thus, although the numerical values of the normalized sensitivities vary slightly with the adopted uncertainty floor, the relative importance of the principal model ingredients remains unchanged. These results demonstrate that the principal conclusions of the sensitivity analysis are robust against reasonable assumptions adopted for the missing experimental uncertainties.

\begin{table*}[htbp]
\centering
\caption{Effect of the adopted minimum relative uncertainty (5\% and 10\%) on the sensitivity analysis for reaction channels involving the dataset of ~\cite{Fukushima1963,Fukushima1965}. The analysis is restricted to these channels because this dataset contains EXFOR entries with zero reported uncertainties.}
\label{tab:uncertainty_floor}
\renewcommand{\arraystretch}{1.2}
\begin{tabular}{lccc|ccc|c}
\toprule
\multirow{2}{*}{Reaction}
& \multicolumn{3}{c|}{5\% uncertainty floor}
& \multicolumn{3}{c|}{10\% uncertainty floor}
& \multirow{2}{*}{Interpretation changed?} \\
\cmidrule(lr){2-4}\cmidrule(lr){5-7}
& $S_{\rm LDM}$ (\%)
& $S_{\alpha\rm OMP}$ (\%)
& $S_{\rm PE}$ (\%)
& $S_{\rm LDM}$ (\%)
& $S_{\alpha\rm OMP}$ (\%)
& $S_{\rm PE}$ (\%)
& \\
\midrule
$^{107}$Ag($\alpha$,2n)$^{109g}$In
& 42.7 & 11.3 & 46.0
& 41.3 & 10.9 & 47.8
& No \\

$^{109}$Ag($\alpha$,3n)$^{110g}$In
& 85.0 & 4.0 & 11.0
& 82.1 & 5.1 & 12.8
& No \\

$^{107}$Ag($\alpha$,n)$^{110m}$In
& 5.1 & 31.7 & 63.2
& 4.9 & 31.3 & 63.8
& No \\
\bottomrule
\end{tabular}

\vspace{2mm}

\end{table*}

\section{Summary}\label{sec5}

In this work, a systematic investigation of 192 combinations of level density models, alpha optical model potentials, and pre-equilibrium models was performed to calculate the cross sections of alpha-induced reactions on silver isotopes leading to the production of the medically relevant indium isotopes $^{108m,109g,110m,110g,111g}$In using the TALYS 2.0 code. The main results are summarized below:

\begin{enumerate}

\item \noindent{$^{107}$Ag($\alpha$,3n)$^{108m}$In:}
This reaction exhibits a strong dominance of the level density model, as reflected by the large normalized sensitivity ($\sim$87.5\%). The effects of the optical model potential and pre-equilibrium mechanisms are comparatively minor. 

\item \noindent{$^{107}$Ag($\alpha$,2n)$^{109g}$In:}
A balanced sensitivity between LDM ($\sim$42.7\%) and PE ($\sim$46.0\%) is observed, with OMP playing a secondary role. The relatively narrow spread in $\chi^2_\nu$ values and limited number of better-performing combinations suggest a well-constrained model space. This indicates a mixed reaction mechanism involving both compound nucleus formation and pre-equilibrium processes. However, this result should be interpreted with caution due to its reliance on a single early experimental data set, highlighting the need for new, precise measurements.

\item \noindent{$^{109}$Ag($\alpha$,3n)$^{110g}$In:}
This reaction is strongly dominated by LDM ($\sim$85\%), with minimal contributions from OMP and PE. Although a relatively broad variation in $\chi^2_\nu$ is observed across model combinations, the reaction mechanism is predominantly governed by compound nucleus formation, underscoring the importance of level density prescriptions.

\item \noindent{$^{107}$Ag($\alpha$,n)$^{110m}$In:}
This reaction is predominantly governed by the pre-equilibrium mechanism ($\sim$63\%), with the $\alpha$-optical model potential making a substantial secondary contribution ($\sim$32\%), while the level-density model plays only a minor role ($\sim$5\%).

\item \noindent{$^{109}$Ag($\alpha$,2n)$^{111g}$In:}
This reaction is predominantly governed by pre-equilibrium processes ($\sim$63\%), with OMP contributing moderately ($\sim$22\%) and LDM playing a relatively minor role ($\sim$15\%). The relatively narrow spread in $\chi^2_\nu$ values suggests stable optimization behavior.

\item The pre-equilibrium sensitivity is found to be channel dependent. While the PE4 option gives the best agreement for one reaction channel, it does not provide consistently superior performance across the remaining channels. Consequently, its apparent preference should be interpreted as a model-dependent local optimum rather than definitive evidence for the underlying reaction mechanism.

\item When EXFOR data show significant scatter across different experimental datasets, the calculated $\chi^2_\nu$ values may partly reflect inconsistencies within the data rather than purely the model’s performance.

\item The variation in performance of different model combinations across nuclides mainly reflects differences in nuclear structure and reaction mechanisms. Key parameters-such as level density, optical potentials, and pre-equilibrium effects-depend on target mass and energy. As a result, a model that works well for one nuclide may not be equally effective for another, especially when structural features like shell effects or deformation vary.

\item The discrepancies between TALYS 2.0 results and TENDL-2023 data stem from the fact that TENDL is an evaluated library derived from systematically optimized and locally adjusted model parameters (e.g., optical model potentials, level densities, and pre-equilibrium contributions), rather than a direct single-run TALYS output. In contrast, the present TALYS 2.0 calculations utilize fixed, predefined model combinations without parameter optimization, leading to the observed deviations.

\item The inability of any of the 192 TALYS-2.0 model combinations to quantitatively reproduce the measured excitation functions indicates that further refinement of the nuclear-reaction models and/or their parameterizations may be required for describing $\alpha$-induced reactions in the $A\approx100$ mass region.

\item Future cross-section measurements for $^{109}$Ag($\alpha$,3n)$^{110g}$In and $^{107}$Ag($\alpha$,n)$^{110m}$In reactions at common projectile energies are essential to enable isomeric ratio calculation which provide a sensitive observable for probing angular momentum population and nuclear level density effects, offering valuable insight into reaction mechanisms.

\end{enumerate}

\vspace{0.3cm}


The sensitivity analysis reveals distinct channel-dependent behaviour of the statistical-model ingredients. The $^{107}$Ag($\alpha$,3n)$^{108m}$In and $^{109}$Ag($\alpha$,3n)$^{110g}$In reaction channels are primarily governed by the level-density model, indicating that compound-nucleus formation and its subsequent statistical decay dominate these multi-neutron emission processes. In contrast, the $^{107}$Ag($\alpha$,2n)$^{109g}$In reaction exhibits nearly comparable sensitivities to the LDM and the pre-equilibrium mechanism, suggesting that both equilibrium and pre-equilibrium emission contribute significantly to the calculated cross sections. The $^{107}$Ag($\alpha$,n)$^{110m}$In and $^{109}$Ag($\alpha$,2n)$^{111g}$In reactions are predominantly governed by the PE mechanism, with the $\alpha$-optical model potential ($\alpha$OMP) providing a secondary contribution and the LDM playing only a minor role. These results demonstrate that the relative importance of the statistical-model ingredients varies considerably among the investigated reaction channels, emphasizing that the dominant reaction mechanism depends on the specific reaction under consideration.

Overall, the observed distribution of the normalized sensitivities, $S_i$, demonstrates that the relative importance of the statistical-model ingredients is strongly reaction-channel dependent. While multi-neutron emission channels are generally controlled by the LDM, reactions involving fewer emitted neutrons exhibit an increased influence of the PE mechanism. These results demonstrate that the optimal statistical-model parameter combination depends on the reaction channel under consideration. Consequently, reliable reproduction of experimental excitation functions requires reaction-specific parameter optimization rather than the use of a single parameter set for all channels.


However, the present $\chi^2$ analysis is based primarily on the reported experimental uncertainties, supplemented by a minimum uncertainty-floor test for datasets containing entries without quoted uncertainties. To provide a qualitative assessment of the model dependence, model-dispersion bands have also been constructed from the complete ensemble of 192 TALYS calculations by taking the minimum and maximum predicted cross sections at each incident alpha energy. These bands illustrate the spread of the calculated excitation functions arising from the investigated variations of the statistical-model parameters, but they do not constitute rigorous theoretical uncertainty or covariance intervals. A full propagation of theoretical model uncertainties, for example using the TASMAN framework, is beyond the scope of the present work. Therefore, the preferred model combinations identified in this study should be interpreted as comparative indicators within the adopted TALYS~2.0 framework rather than as definitive globally optimized solutions. Moreover, since TENDL-2023 is generated using TALYS-based calculations with adjusted model parameters, the present comparison may also be viewed as an assessment of the sensitivity of the default TALYS~2.0 statistical-model options relative to the evaluated TENDL predictions.

A sensitivity test may be performed as a future extension of this work by re-running the optimization for reactions with multiple data sets by including and excluding each set in turn to check whether the model combination with comparatively better agreement with the experimental data changes.

\section*{Acknowledgements}

The author thanks ICFAI University Tripura for all the helpful support in carrying out the research work.

\section*{Data Availability Statement}

The experimental data in the present work has been taken from the Experimental Nuclear Reaction Data (EXFOR) library of the International Network of Nuclear Reaction Data Centers (NRDC) (DOI: 10.1016/j.nds.2014.07.065)																	
\clearpage
\appendix
\renewcommand{\thetable}{A\arabic{table}}
\setcounter{table}{0}

\renewcommand{\thefigure}{A\arabic{figure}}
\setcounter{figure}{0}

\section{\label{sec:1}Supplementary results of reduced $\chi^2$ analysis for the different combination of theoretical models}

For completeness, the reduced $\chi^2$ values corresponding to all 192 combinations of LDMs, $\alpha$OMPs, and PEs have been computed for each investigated reaction; the results are presented in five separate tables below. In all tables, the reduced $\chi^2$ values corresponding to the statistically equivalent better-performing combinations are highlighted in bold.

\begin{table}[H]
\centering
\caption{Reduced $\chi^2$ values for 192 possible combinations of LDMs, $\alpha$OMPs and PEs for $^{107}$Ag($\alpha$,3n)$^{108m}$In. All combinations yielding reduced $\chi^2$ values within 5\% of the minimum are highlighted in bold.}
\begin{tabular}{ccccccccc}
\hline
with PE1	&	$\alpha$OMP1	&	$\alpha$OMP2	&	$\alpha$OMP3	&	$\alpha$OMP4	&	$\alpha$OMP5	&	$\alpha$OMP6	&	$\alpha$OMP7	&	$\alpha$OMP8	\\
\hline																	
LDM1	&	198	&	210	&	196	&	197	&	197	&	224	&	242	&	234	\\
LDM2	&	121	&	128	&	119	&	120	&	119	&	135	&	144	&	140	\\
LDM3	&	101	&	105	&	100	&	101	&	101	&	109	&	114	&	111	\\
LDM4	&	58	&	59	&	58	&	58	&	58	&	59	&	59	&	59	\\
LDM5	&	52	&	53	&	51	&	52	&	51	&	53	&	51	&	51	\\
LDM6	&	\textbf{49}	&	\textbf{50}	&	\textbf{49}	&	\textbf{49}	&	\textbf{49}	&	\textbf{50}	&	\textbf{50}	&	\textbf{50}	\\
\hline																		
with PE2	&	$\alpha$OMP1	&	$\alpha$OMP2	&	$\alpha$OMP3	&	$\alpha$OMP4	&	$\alpha$OMP5	&	$\alpha$OMP6	&	$\alpha$OMP7	&	$\alpha$OMP8	\\
\hline																	
LDM1	&	199	&	211	&	196	&	198	&	198	&	225	&	243	&	235	\\
LDM2	&	127	&	128	&	119	&	120	&	120	&	136	&	145	&	140	\\
LDM3	&	101	&	105	&	100	&	101	&	101	&	109	&	114	&	111	\\
LDM4	&	58	&	59	&	58	&	58	&	58	&	60	&	59	&	59	\\
LDM5	&	52	&	53	&	51	&	52	&	51	&	53	&	51	&	51	\\
LDM6	&	\textbf{49}	&	\textbf{50}	&	\textbf{49}	&	\textbf{49}	&	\textbf{49}	&	\textbf{50}	&	\textbf{50}	&	\textbf{50}	\\
\hline																		
with PE3	&	$\alpha$OMP1	&	$\alpha$OMP2	&	$\alpha$OMP3	&	$\alpha$OMP4	&	$\alpha$OMP5	&	$\alpha$OMP6	&	$\alpha$OMP7	&	$\alpha$OMP8	\\
\hline																	
LDM1	&	187	&	198	&	184	&	186	&	185	&	211	&	229	&	221	\\
LDM2	&	114	&	120	&	112	&	113	&	113	&	128	&	136	&	132	\\
LDM3	&	97	&	100	&	96	&	96	&	96	&	104	&	108	&	106	\\
LDM4	&	56	&	57	&	56	&	56	&	56	&	58	&	57	&	57	\\
LDM5	&	52	&	53	&	51	&	52	&	51	&	53	&	52	&	52	\\
LDM6	&	\textbf{48}	&	\textbf{49}	&	\textbf{48}	&	\textbf{48}	&	\textbf{48}	&	\textbf{49}	&	\textbf{49}	&	\textbf{49}	\\
\hline																		
with PE4	&	$\alpha$OMP1	&	$\alpha$OMP2	&	$\alpha$OMP3	&	$\alpha$OMP4	&	$\alpha$OMP5	&	$\alpha$OMP6	&	$\alpha$OMP7	&	$\alpha$OMP8	\\
\hline																	
LDM1	&	205	&	215	&	202	&	204	&	203	&	228	&	249	&	241	\\
LDM2	&	126	&	132	&	124	&	125	&	125	&	140	&	149	&	144	\\
LDM3	&	105	&	108	&	104	&	104	&	104	&	113	&	117	&	114	\\
LDM4	&	62	&	64	&	62	&	62	&	62	&	65	&	63	&	62	\\
LDM5	&	55	&	57	&	55	&	55	&	55	&	57	&	54	&	53	\\
LDM6	&	53	&	55	&	53	&	53	&	53	&	55	&	54	&	53	\\
\hline
\label{tab1}
\end{tabular}
\end{table}

\begin{table}[H]
\centering
\caption{Same as Table~\ref{tab1}, for $^{107}$Ag($\alpha$,2n)$^{109g}$In reaction}
\begin{tabular}{ccccccccc}
\hline
with PE1	&	$\alpha$OMP1	&	$\alpha$OMP2	&	$\alpha$OMP3	&	$\alpha$OMP4	&	$\alpha$OMP5	&	$\alpha$OMP6	&	$\alpha$OMP7	&	$\alpha$OMP8	\\
\hline																	
LDM1	&	48	&	45	&	53	&	49	&	49	&	43	&	45	&	48	\\
LDM2	&	62	&	59	&	64	&	63	&	63	&	57	&	45	&	60	\\
LDM3	&	85	&	81	&	86	&	85	&	85	&	79	&	80	&	83	\\
LDM4	&	63	&	60	&	65	&	64	&	64	&	57	&	56	&	57	\\
LDM5	&	31	&	28	&	32	&	31	&	32	&	27	&	35	&	36	\\
LDM6	&	32	&	29	&	34	&	33	&	33	&	27	&	29	&	30	\\
\hline																	
																	
with PE2	&	$\alpha$OMP1	&	$\alpha$OMP2	&	$\alpha$OMP3	&	$\alpha$OMP4	&	$\alpha$OMP5	&	$\alpha$OMP6	&	$\alpha$OMP7	&	$\alpha$OMP8	\\
\hline																	
LDM1	&	48	&	46	&	49	&	49	&	49	&	43	&	45	&	48	\\
LDM2	&	62	&	59	&	64	&	63	&	63	&	57	&	49	&	60	\\
LDM3	&	85	&	81	&	86	&	85	&	85	&	79	&	80	&	83	\\
LDM4	&	63	&	61	&	66	&	64	&	64	&	57	&	56	&	57	\\
LDM5	&	31	&	28	&	32	&	31	&	32	&	27	&	35	&	36	\\
LDM6	&	32	&	29	&	34	&	33	&	33	&	27	&	29	&	30	\\
\hline																	
																	
with PE3	&	$\alpha$OMP1	&	$\alpha$OMP2	&	$\alpha$OMP3	&	$\alpha$OMP4	&	$\alpha$OMP5	&	$\alpha$OMP6	&	$\alpha$OMP7	&	$\alpha$OMP8	\\
\hline																	
LDM1	&	44	&	41	&	45	&	44	&	44	&	39	&	41	&	44	\\
LDM2	&	59	&	56	&	60	&	59	&	60	&	53	&	54	&	56	\\
LDM3	&	83	&	80	&	85	&	84	&	84	&	77	&	78	&	81	\\
LDM4	&	58	&	55	&	60	&	59	&	59	&	52	&	50	&	52	\\
LDM5	&	29	&	27	&	31	&	30	&	30	&	\textbf{26}	&	33	&	35	\\
LDM6	&	30	&	28	&	32	&	31	&	31	&	\textbf{25}	&	27	&	28	\\
\hline																	
																	
with PE4	&	$\alpha$OMP1	&	$\alpha$OMP2	&	$\alpha$OMP3	&	$\alpha$OMP4	&	$\alpha$OMP5	&	$\alpha$OMP6	&	$\alpha$OMP7	&	$\alpha$OMP8	\\
\hline																	
LDM1	&	94	&	134	&	95	&	94	&	94	&	138	&	71	&	69	\\
LDM2	&	116	&	154	&	118	&	117	&	117	&	156	&	97	&	94	\\
LDM3	&	128	&	166	&	130	&	129	&	129	&	169	&	107	&	105	\\
LDM4	&	122	&	156	&	124	&	123	&	123	&	158	&	108	&	104	\\
LDM5	&	68	&	109	&	70	&	69	&	69	&	112	&	50	&	47	\\
LDM6	&	81	&	117	&	84	&	82	&	82	&	120	&	65	&	62	\\
\hline																	
\label{tab2}
\end{tabular}
\end{table}

\begin{table}[H]
\centering
\caption{Same as Table~\ref{tab1}, for $^{107}$Ag($\alpha$,n)$^{110m}$In}
\begin{tabular}{ccccccccc}
\hline
with PE1	&	$\alpha$OMP1	&	$\alpha$OMP2	&	$\alpha$OMP3	&	$\alpha$OMP4	&	$\alpha$OMP5	&	$\alpha$OMP6	&	$\alpha$OMP7	&	$\alpha$OMP8	\\
\hline																	
LDM1	&	142	&	145	&	141	&	142	&	142	&	160	&	781	&	263	\\
LDM2	&	126	&	127	&	124	&	125	&	125	&	139	&	736	&	227	\\
LDM3	&	\textbf{61}	&	\textbf{60}	&	\textbf{61}	&	\textbf{61}	&	\textbf{61}	&	71	&	620	&	140	\\
LDM4	&	115	&	115	&	114	&	114	&	114	&	124	&	687	&	206	\\
LDM5	&	101	&	100	&	100	&	101	&	101	&	108	&	651	&	184	\\
LDM6	&	216	&	225	&	210	&	213	&	213	&	253	&	823	&	349	\\
\hline																	
																	
with PE2	&	$\alpha$OMP1	&	$\alpha$OMP2	&	$\alpha$OMP3	&	$\alpha$OMP4	&	$\alpha$OMP5	&	$\alpha$OMP6	&	$\alpha$OMP7	&	$\alpha$OMP8	\\
\hline	
LDM1	&	142	&	144	&	140	&	141	&	141	&	160	&	780	&	262	\\
LDM2	&	125	&	126	&	124	&	125	&	125	&	139	&	735	&	226	\\
LDM3	&	\textbf{61}	&	\textbf{60}	&	\textbf{61}	&	\textbf{61}	&	\textbf{61}	&	71	&	620	&	140	\\
LDM4	&	114	&	114	&	113	&	114	&	114	&	124	&	687	&	206	\\
LDM5	&	101	&	100	&	100	&	101	&	101	&	108	&	650	&	183	\\
LDM6	&	216	&	225	&	210	&	213	&	212	&	253	&	823	&	349	\\
\hline

with PE3	&	$\alpha$OMP1	&	$\alpha$OMP2	&	$\alpha$OMP3	&	$\alpha$OMP4	&	$\alpha$OMP5	&	$\alpha$OMP6	&	$\alpha$OMP7	&	$\alpha$OMP8	\\
\hline																	
LDM1	&	147	&	150	&	146	&	147	&	147	&	166	&	787	&	269	\\
LDM2	&	129	&	130	&	128	&	129	&	129	&	143	&	740	&	231	\\
LDM3	&	\textbf{61}	&	\textbf{60}	&	\textbf{61}	&	\textbf{61}	&	\textbf{61}	&	71	&	620	&	140	\\
LDM4	&	118	&	118	&	117	&	118	&	118	&	128	&	692	&	210	\\
LDM5	&	104	&	103	&	103	&	103	&	103	&	111	&	654	&	187	\\
LDM6	&	219	&	228	&	213	&	216	&	215	&	256	&	826	&	352	\\
\hline																	
																	
with PE4	&	$\alpha$OMP1	&	$\alpha$OMP2	&	$\alpha$OMP3	&	$\alpha$OMP4	&	$\alpha$OMP5	&	$\alpha$OMP6	&	$\alpha$OMP7	&	$\alpha$OMP8	\\
\hline																	
LDM1	&	4292	&	13180	&	4288	&	4290	&	4289	&	16942	&	5042	&	3551	\\
LDM2	&	4132	&	12930	&	4127	&	4130	&	4129	&	16666	&	4841	&	3389	\\	
LDM3	&	3011	&	10820	&	3008	&	3009	&	3009	&	14309	&	3688	&	2494	\\	
LDM4	&	4146	&	12982	&	4142	&	4144	&	4143	&	16733	&	4799	&	3375	\\	
LDM5	&	4016	&	12802	&	4014	&	4015	&	4015	&	16549	&	4655	&	3254	\\	
LDM6	&	4294	&	13131	&	4283	&	4289	&	4288	&	16873	&	5048	&	3613	\\	
\hline																	
\label{tab3}
\end{tabular}
\end{table}

\begin{table}[H]
\centering
\caption{Same as Table~\ref{tab1}, for $^{109}$Ag($\alpha$,3n)$^{110g}$In}
\begin{tabular}{ccccccccc}
\hline
with PE1	&	$\alpha$OMP1	&	$\alpha$OMP2	&	$\alpha$OMP3	&	$\alpha$OMP4	&	$\alpha$OMP5	&	$\alpha$OMP6	&	$\alpha$OMP7	&	$\alpha$OMP8	\\
\hline																	
LDM1	&	40	&	39	&	41	&	40	&	41	&	37	&	\textbf{35}	&	\textbf{36}	\\
LDM2	&	53	&	52	&	54	&	54	&	54	&	50	&	48	&	49	\\
LDM3	&	117	&	116	&	117	&	117	&	117	&	115	&	115	&	115	\\
LDM4	&	60	&	59	&	61	&	61	&	61	&	57	&	58	&	58	\\
LDM5	&	71	&	71	&	74	&	72	&	73	&	69	&	77	&	75	\\
LDM6	&	81	&	80	&	82	&	81	&	82	&	79	&	80	&	80	\\
\hline

with PE2	&	$\alpha$OMP1	&	$\alpha$OMP2	&	$\alpha$OMP3	&	$\alpha$OMP4	&	$\alpha$OMP5	&	$\alpha$OMP6	&	$\alpha$OMP7	&	$\alpha$OMP8	\\
\hline																	
LDM1	&	40	&	39	&	41	&	40	&	40	&	37	&	\textbf{35}	&	\textbf{36}	\\
LDM2	&	53	&	52	&	54	&	54	&	54	&	50	&	48	&	49	\\
LDM3	&	117	&	116	&	117	&	117	&	117	&	115	&	115	&	115	\\
LDM4	&	60	&	59	&	61	&	61	&	61	&	57	&	58	&	58	\\
LDM5	&	81	&	80	&	82	&	81	&	82	&	79	&	80	&	80	\\
LDM6	&	71	&	71	&	74	&	72	&	73	&	69	&	77	&	75	\\
\hline																	
																	
with PE3	&	$\alpha$OMP1	&	$\alpha$OMP2	&	$\alpha$OMP3	&	$\alpha$OMP4	&	$\alpha$OMP5	&	$\alpha$OMP6	&	$\alpha$OMP7	&	$\alpha$OMP8	\\
\hline																	
LDM1	&	40	&	39	&	41	&	41	&	41	&	37	&	\textbf{34}	&	\textbf{35}	\\
LDM2	&	53	&	52	&	54	&	54	&	54	&	50	&	48	&	49	\\
LDM3	&	115	&	114	&	115	&	115	&	115	&	113	&	113	&	113	\\
LDM4	&	60	&	59	&	61	&	61	&	61	&	57	&	57	&	57	\\
LDM5	&	70	&	70	&	73	&	71	&	71	&	68	&	76	&	74	\\
LDM6	&	80	&	79	&	81	&	81	&	81	&	78	&	78	&	79	\\
\hline																	
																	
with PE4	&	$\alpha$OMP1	&	$\alpha$OMP2	&	$\alpha$OMP3	&	$\alpha$OMP4	&	$\alpha$OMP5	&	$\alpha$OMP6	&	$\alpha$OMP7	&	$\alpha$OMP8	\\
\hline																	
LDM1	&	47	&	52	&	47	&	47	&	47	&	52	&	40	&	40	\\
LDM2	&	60	&	66	&	61	&	60	&	61	&	60	&	54	&	54	\\
LDM3	&	121	&	124	&	122	&	122	&	122	&	124	&	119	&	119	\\
LDM4	&	66	&	72	&	67	&	67	&	67	&	69	&	62	&	63	\\
LDM5	&	78	&	84	&	81	&	79	&	80	&	84	&	84	&	81	\\
LDM6	&	87	&	91	&	88	&	88	&	88	&	91	&	86	&	85	\\
\hline											
\label{tab4}
\end{tabular}
\end{table}

\begin{table}[H]
\centering
\caption{Same as Table~\ref{tab1}, for $^{109}$Ag($\alpha$,2n)$^{111g}$In}
\begin{tabular}{ccccccccc}
\hline	
with PE1	&	$\alpha$OMP1	&	$\alpha$OMP2	&	$\alpha$OMP3	&	$\alpha$OMP4	&	$\alpha$OMP5	&	$\alpha$OMP6	&	$\alpha$OMP7	&	$\alpha$OMP8	\\
\hline																	
LDM1	&	171	&	181	&	169	&	169	&	169	&	195	&	249	&	237	\\
LDM2	&	150	&	159	&	147	&	148	&	148	&	171	&	214	&	204	\\
LDM3	&	153	&	162	&	152	&	152	&	152	&	173	&	216	&	207	\\
LDM4	&	146	&	154	&	144	&	145	&	144	&	166	&	203	&	196	\\
LDM5	&	173	&	183	&	171	&	171	&	171	&	198	&	250	&	239	\\
LDM6	&	134	&	141	&	130	&	132	&	131	&	152	&	179	&	174	\\
\hline

with PE2	&	$\alpha$OMP1	&	$\alpha$OMP2	&	$\alpha$OMP3	&	$\alpha$OMP4	&	$\alpha$OMP5	&	$\alpha$OMP6	&	$\alpha$OMP7	&	$\alpha$OMP8	\\
\hline																	
LDM1	&	171	&	181	&	169	&	169	&	169	&	196	&	250	&	237	\\
LDM2	&	150	&	159	&	147	&	148	&	148	&	171	&	214	&	205	\\
LDM3	&	154	&	162	&	152	&	152	&	152	&	174	&	216	&	207	\\
LDM4	&	146	&	154	&	144	&	145	&	145	&	166	&	203	&	196	\\
LDM5	&	173	&	184	&	171	&	172	&	171	&	198	&	251	&	239	\\
LDM6	&	134	&	141	&	130	&	132	&	171	&	152	&	179	&	174	\\
\hline																	
																	
with PE3	&	$\alpha$OMP1	&	$\alpha$OMP2	&	$\alpha$OMP3	&	$\alpha$OMP4	&	$\alpha$OMP5	&	$\alpha$OMP6	&	$\alpha$OMP7	&	$\alpha$OMP8	\\
\hline																	
LDM1	&	170	&	180	&	168	&	169	&	168	&	195	&	249	&	237	\\
LDM2	&	149	&	158	&	147	&	148	&	147	&	171	&	214	&	204	\\
LDM3	&	152	&	160	&	150	&	150	&	150	&	171	&	214	&	204	\\
LDM4	&	147	&	155	&	144	&	145	&	145	&	167	&	204	&	197	\\
LDM5	&	174	&	184	&	171	&	172	&	172	&	199	&	251	&	240	\\
LDM6	&	138	&	145	&	134	&	136	&	135	&	156	&	184	&	179	\\
\hline

with PE4	&	$\alpha$OMP1	&	$\alpha$OMP2	&	$\alpha$OMP3	&	$\alpha$OMP4	&	$\alpha$OMP5	&	$\alpha$OMP6	&	$\alpha$OMP7	&	$\alpha$OMP8	\\
\hline																	
LDM1	&	30	&	33	&	30	&	30	&	30	&	34	&	28	&	27	\\
LDM2	&	30	&	34	&	31	&	30	&	31	&	35	&	28	&	28	\\
LDM3	&	41	&	43	&	41	&	41	&	41	&	43	&	39	&	39	\\
LDM4	&	27	&	31	&	27	&	27	&	27	&	31	&	25	&	24	\\
LDM5	&	24	&	28	&	25	&	24	&	24	&	29	&	\textbf{22}	&	\textbf{22}	\\
LDM6	&	24	&	28	&	24	&	24	&	24	&	28	&	\textbf{21}	&	\textbf{21}	\\
\hline																											
\label{tab5}
\end{tabular}
\end{table}

\printcredits

\bibliographystyle{cas-model2-names}
\clearpage
\bibliography{cas-refs}


\end{document}